\pdfoutput=1

\documentclass[11pt]{article}

\usepackage[preprint]{acl}

\usepackage{times}
\usepackage{latexsym}
\usepackage{amsmath}
\usepackage{graphicx}
\usepackage[T1]{fontenc}

\usepackage[utf8]{inputenc}

\usepackage{microtype}

\usepackage{inconsolata}

\usepackage{graphicx}
\usepackage{booktabs}
\usepackage{multirow}
\usepackage{bbding}
\usepackage{utfsym}
\usepackage{enumitem}

\title{UniCodec: Unified Audio Codec with Single Domain-Adaptive Codebook}

\author{
 \textbf{Yidi Jiang\textsuperscript{1,2}},
 \textbf{Qian Chen\textsuperscript{2}\thanks{Corresponding author.}},
 \textbf{Shengpeng Ji\textsuperscript{2}},
 \textbf{Yu Xi\textsuperscript{3}},
\textbf{Wen Wang\textsuperscript{2}}, \\
 \textbf{Chong Zhang\textsuperscript{2}},
 \textbf{Xianghu Yue\textsuperscript{1}},
 \textbf{Shiliang Zhang\textsuperscript{2}},
 \textbf{Haizhou Li\textsuperscript{4,5,1}}
\\
 \textsuperscript{1}National University of Singapore, Singapore~~~~
 \textsuperscript{2}Tongyi Speech Lab \\
 \textsuperscript{3}Shanghai Jiao Tong University~~~
 \textsuperscript{4}Shenzhen Research Institute of Big Data, Shenzhen \\
 \textsuperscript{5}School of Data Science, The Chinese University of Hong Kong, Shenzhen
}

\begin{document}
\maketitle
\begin{abstract}
    The emergence of audio language models is empowered by neural audio codecs, which establish critical mappings between continuous waveforms and discrete tokens compatible with language model paradigms. 
    The evolutionary trends from multi-layer residual vector quantizer to single-layer quantizer are beneficial for language-autoregressive decoding.
    However, the capability to handle multi-domain audio signals through a single codebook remains constrained by inter-domain distribution discrepancies.
    In this work, we introduce \textbf{UniCodec}, a unified audio codec with a single codebook to support multi-domain audio data, including \textit{speech}, \textit{music}, and \textit{sound}. To achieve this, we propose a partitioned domain-adaptive codebook method and domain Mixture-of-Experts strategy to capture the distinct characteristics of each audio domain. Furthermore, to enrich the semantic density of the codec without auxiliary modules, we propose a self-supervised mask prediction modeling approach. 
    Comprehensive objective and subjective evaluations demonstrate that UniCodec achieves excellent audio reconstruction performance across the three audio domains, outperforming existing unified neural codecs with a single codebook, and even surpasses state-of-the-art \textit{domain-specific} codecs on both acoustic and semantic representation capabilities\footnote{https://github.com/Jiang-Yidi/UniCodec}.

\end{abstract}

\section{Introduction}

Many recent developments of speech language models (SLMs) \cite{bai2023qwen,defossez2024moshi,peng2024survey,ji2024wavchat} integrate the speech modality with text-based large language models (LLMs) and have led to significant advancements in speech understanding and generation tasks. This paradigm relies on discrete acoustic codec models, which convert high-rate speech signals into a finite set of discrete speech tokens, bridging the gap between continuous speech signals and discrete-token-based language models, thus enabling speech applications powered by LLMs.


Most existing neural audio codecs (NACs)~\cite{taslp2021-soundstream,nips2023-dac-rvqgan,ji2024language,tmlr2023-encodec,defossez2024moshi} employ a \textit{multi-layer} Residual Vector Quantizer (RVQ), where each quantizer operates on the residual of the previous quantizer. This RVQ structure generates multiple parallel hierarchical token streams for downstream language models to decode, hence it increases the complexity and the generation latency of SLMs~\cite{xie2024mini,xie2024mini2,defossez2024moshi}. 
To address this problem, several recent works, including WavTokenizer~\cite{ji2024wavtokenizer}, Single-Codec~\cite{interspeech2024-singlecodec}, and BigCodec~\cite{arxiv2024-bigcodec}, focus on developing \textit{single-layer} quantizer to streamline the process. Integrating a single-layer quantizer with LLMs facilitates rapid extraction of speech features on input audio while significantly reducing the burden of autoregressive modeling. These works demonstrate that using a single VQ to discretize speech could achieve competitive performance in both audio reconstruction and generation tasks. Therefore, our work follows this trend and focuses on developing high-performing single-layer quantizer codec.

\begin{table*}[!t]
    \centering
    \small
   \caption{Comparison of recent codec models based on single codebook, compatibility with speech, music, and sound
   domains, and whether they use \textit{separate} models for different domains or a \textit{unified} model.}
    \begin{tabular}{c|cccc}
      \toprule
      Model & Single Codebook & Speech & Music\&Sound & Separate/Unified model \\
      \midrule
      DAC~\cite{nips2023-dac-rvqgan}& $\usym{2718}$ & $\usym{2714}$ & $\usym{2714}$ & Unified \\
      Encodec~\cite{tmlr2023-encodec} & $\usym{2718}$ & $\usym{2714}$ & $\usym{2714}$ & Unified \\
      Mimi~\cite{defossez2024moshi} & $\usym{2718}$ & $\usym{2714}$ & $\usym{2714}$ & Unified \\
      SemantiCodec~\cite{liu2024semanticodec} & $\usym{2718}$ & $\usym{2714}$ & $\usym{2714}$ & Unified  \\
      SpeechTokenizer~\cite{zhang2023speechtokenizer} & $\usym{2718}$ & $\usym{2714}$ & $\usym{2718}$ & - \\
      BigCodec~\cite{arxiv2024-bigcodec} & $\usym{2714}$ & $\usym{2714}$ & $\usym{2718}$ & - \\
      TAAE~\cite{arxiv2024-taae} & $\usym{2714}$ & $\usym{2714}$ & $\usym{2718}$ & -\\
      Wavtokenizer~\cite{ji2024wavtokenizer} & $\usym{2714}$ & $\usym{2714}$ & $\usym{2714}$ & Separate\&Unified  \\
      \textbf{UniCodec} & $\usym{2714}$ & $\usym{2714}$ & $\usym{2714}$ & \textbf{Unified} \\
      \bottomrule
    \end{tabular}
    \label{tab:overview}
\end{table*}

An ideal codec should be able to perform well across various audio domains, such as speech, music, and sound, with distinct domain characteristics. Prior RVQ-based neural audio codecs using \textit{multi-layer RVQ and hence multi-codebooks}, such as DAC~\cite{nips2023-dac-rvqgan} and Encodec~\cite{tmlr2023-encodec}, exhibit strong reconstruction capabilities for speech, music, and sound.
However, previous studies such as Wavtokenizer~\cite{ji2024wavtokenizer} show that using a \textit{unified single-codebook codec} for speech, music, and sound still poses a great challenge: The unified codec suffers from notable performance degradation compared to domain-specific codec models, since the substantial distribution discrepancies between these domains make it difficult to effectively capture their distinct characteristics with a single codebook.
To tackle this challenge, in this work,  we develop a \textbf{unified audio codec with a single codebook, designed to support multiple audio domains—including speech, music, and sound—while achieving both low bitrate and high acoustic reconstruction quality}.

In addition to powerful acoustic reconstruction capabilities, strong semantic representation capabilities (that is, encapsulating rich semantic information) of NACs are crucial for effective integration of NACs with LLMs, since strong semantic capabilities can ease understanding of audio content and facilitate generation of semantically reasonable audio. There are two main challenges in enriching the semantic representations of NACs. (1) There is an inherent trade-off between semantic richness and reconstruction performance, since
semantic features provide a higher-level, more abstract understanding, while reconstruction features emphasize fine-grained details of audio. 
(2) The majority of existing works enrich semantic capabilities through distillation from additional pretrained speech semantic encoders~\cite{zhang2023speechtokenizer, defossez2024moshi}, separate semantic codebooks~\cite{liu2024semanticodec}, or auxiliary semantic modules~\cite{xcodec}. However, methods using an additional pretrained semantic encoder are constrained by reliance on a pretrained speech encoder, are less elegant and not fully adaptable, and difficult to support unified modeling of speech, music, and sound.  Moreover, an auxiliary semantic module introduces additional computation cost and degrades the efficiency of feature extraction.
Since both reconstruction quality and efficiency are critical for NACs, we explore a more elegant approach by \textbf{directly learning semantic information through the codec itself, without additional modules, while preserving high reconstruction ability}.

Our contributions can be summarized as follows:
\begin{itemize}[leftmargin=*,noitemsep]
    \item We introduce UniCodec, a unified audio codec with a single quantizer, designed to support various audio types, including speech, music, and sound, with a single codebook. To achieve this, we propose a partitioned domain-adaptive codebook method based on domain Mixture-of-Experts (MoE) strategy to effectively capture the distinct characteristics of each audio domain. 
    \item We propose a self-supervised, masked modeling approach to enrich semantic information without extra modules.
    \item Comprehensive objective and subjective evaluations show that UniCodec achieves better reconstruction and semantic performance compared to existing unified codecs with a single codebook, and even outperforms domain-specific codecs.
\end{itemize}

\section{Related Work}
\paragraph{Neural Audio Codecs}
Neural Audio Codecs (NACs) aim to compress audio signals into highly compressed discrete tokens while preserving high reconstruction quality. The predominant paradigm of NACs utilizes the Vector Quantized Variational Autoencoder (VQ-VAE)~\cite{nips2017-vqvae,icassp2019-vqvaeaudio} architecture, where an encoder transforms the audio signal into a latent representation, a quantizer discretizes this representation, and a decoder reconstructs the signal. SoundStream~\cite{taslp2021-soundstream} enhances this approach by incorporating Residual Vector Quantization (RVQ), and improves both modeling and reconstruction capabilities for NACs. Encodec~\cite{tmlr2023-encodec} further refines SoundStream by introducing multi-scale discriminators and a loss-balancing strategy to optimize reconstruction performance. Numerous works such as DAC~(also named RVQGAN)~\cite{nips2023-dac-rvqgan} and Mimi~\cite{defossez2024moshi} continue enhancing RVQ-based NACs. While multi-codebook residual modeling boosts reconstruction quality, it complicates the autoregressive process in SLMs and suffers from unacceptable latency. In contrast, single-layer quantizer codecs, such as Single-Codec~\cite{interspeech2024-singlecodec}, WavTokenizer~\cite{ji2024wavtokenizer}, BigCodec~\cite{arxiv2024-bigcodec}, and TAAE~\cite{arxiv2024-taae}, show promising potentials due to their ability to seamlessly integrate into SLMs with low latency and reduced computational overhead. However, there is still much room to improve the performance of single-layer low-bitrate codecs; hence, this work focuses on enhancing single-layer low-bitrate codecs.

\paragraph{Unified Audio Signal Modeling}
A unified NAC capable of processing various audio types, such as speech, music, and sound, will be greatly beneficial for constructing universal audio language models (ALMs) that are generalizable to various audio types. RVQ-based audio codec models, such as SoundStream~\cite{taslp2021-soundstream}, Encodec~\cite{tmlr2023-encodec}, and DAC~\cite{nips2023-dac-rvqgan}, are trained on a combination of speech, music, and sound datasets. While these codecs achieve high reconstruction quality, their performance significantly degrades in low-bitrate scenarios, particularly when restricted to the first codebook. Although existing single-layer codecs~\cite{ji2024wavtokenizer} perform well in one or two audio domains, they struggle to simultaneously maintain superior performance on speech, music, and sound domains while operating at a low bitrate. 


\paragraph{Semantic Audio Representation Learning}
Discrete tokens compressed by acoustic NACs lack high-level semantic information, which is essential for effective SLMs. To address this issue, models such as SpeechTokenizer~\cite{zhang2023speechtokenizer} and Mimi~\cite{defossez2024moshi} leverage self-supervised-learning (SSL) based speech representation models to distill semantic information into the first-layer codebook. XCodec~\cite{xcodec} concatenates acoustic tokens with semantic tokens produced by SSL models before the RVQ stage and introduces a semantic reconstruction loss. FunCodec~\cite{icassp2024-funcodec} offers various methods to integrate SSL-based semantic tokens with RVQ-based acoustic tokens. However, these approaches rely on SSL encoders, which complicate the training process and constrain the semantic capabilities of NACs. SemantiCodec~\cite{liu2024semanticodec} combines quantized semantic tokens with acoustic tokens and introduces a diffusion process to enhance reconstruction quality, but the diffusion process introduces additionally training cost. In contrast, UniCodec requires neither additional SSL encoders nor complex diffusion process, hence simplifying the training process while encapsulating rich semantic information.

\section{Methodology}
\label{sec:method}
\begin{figure}[t!]
  \centering
  \includegraphics[width=\linewidth]{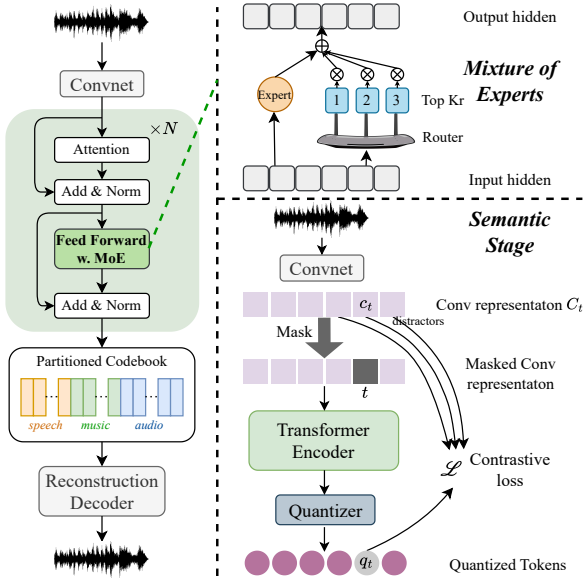}
  \caption{Left: Overview of the proposed UniCodec. Upper-right: the domain MoE encoder structure. Lower-right: the semantic training stage.} 
  \label{fig:overviw}
\end{figure}

UniCodec is built upon the highly competitive single-layer encoder-VQ-decoder codec, Wavtokenizer~\cite{ji2024wavtokenizer}. 
The left part of Figure~\ref{fig:overviw} provides an overview of the architecture of UniCodec, which comprises three modules: an encoder that processes the input audio to generate a latent feature representation, a quantizer that discretizes the feature into tokens through a single codebook, and a decoder that reconstructs the audio signal from the compressed, discrete tokens.
We first make the following enhancements over Wavtokenizer (Section~\ref{subsec:encoder}). We enhance the encoder by incorporating transformer layers, which can better capture and represent complex patterns. We also enhance the codebook utilization rate to maximize the use of codebook and improve efficiency.
More importantly, to build a unified tokenizer capable of supporting multi-domain audio reconstruction, we propose two novel strategies: a partitioned domain-adaptive codebook (Section~\ref{subsec:codebook}), and a domain mixture-of-experts (MoE) encoder structure~(Section~\ref{subsec:moe}), which is detailed in the upper-right part of Figure~\ref{fig:overviw}.
UniCodec is trained end-to-end through two stages. In the first acoustic training stage, the model is trained by optimizing a reconstruction loss applied over both time and frequency domains, along with a perceptual loss using discriminators operating at different resolutions, the same as Wavtokenizer. In the following semantic training stage (Section~\ref{subsec:semantic_stage}), which is depicted in the lower-right part of Figure~\ref{fig:overviw}), a contrastive loss is added into the training objective.

\subsection{Enhanced Encoder and Quantizer}
\label{subsec:encoder}
The encoder of Wavtokenizer~\cite{ji2024wavtokenizer} consists of convolutional blocks followed by a two-layer LSTM and a final 1D convolution layer, which limits its capacity for effective feature extraction. To enhance the ability to encode audio into compact representations while ensuring high-quality audio reconstruction, inspired by Mimi Codec in Moshi~\cite{defossez2024moshi}, we replace the LSTM sequence modeling in the encoder with a contextual Transformer architecture following the convolutional blocks. Consistent with Mimi, the Transformer consists of 8 layers, 8 attention heads, RoPE position encodings, GELU activations~\cite{hendrycks2016bridging}, with a hidden size of 512 and an MLP dimension of 2048.

Scaling the training data to cover multiple audio domains necessitates scaling the codebook concurrently, which introduces the challenge of optimizing codebook utilization during the vector quantization process. To improve codebook utilization and improve efficiency, we adopt the SimVQ algorithm~\cite{simvq}, which effectively and efficiently mitigates the issue of representation collapse in vector-quantized model by using a simple linear layer.

\subsection{Domain-adaptive Codebook}
\label{subsec:codebook}
To achieve seamless integration of data from three distinct domains—speech, music, and sound—into a unified audio tokenizer, we propose a novel partitioned domain-adaptive codebook. In this framework, the codebook is divided into three specialized regions: the first region, spanning indices 0 to 4095, is dedicated to the speech domain; the second, from 4096 to 8191, is for the music domain; and the remaining indices from 8191 to 16383 are allocated for the sound domain.
This design is inspired by the hypothesis in Semanticodec~\cite{liu2024semanticodec} that general sound tends to encompass a broader range of sounds than speech and music, hence we allocate a larger region for sound. During the training process, the model only updates the codebook entries corresponding to the domain of the input sample, ensuring that domain-specific features are accurately captured and learned. 
This partitioned codebook approach facilitates the construction of a unified audio tokenizer that can effectively handle the unique characteristics of each domain, providing a flexible solution for multi-domain audio representation.
The ablation experimental results in Table~\ref{tab:ablation} of Section~\ref{subsec:ablation} validate this strategy achieves performance improvement when scaling up the amount of training data covering different audio types and also codebook size.

\subsection{Domain MoE}
\label{subsec:moe}
For training the codec on data from multiple audio domains, we employ a domain Mixture-of-Experts (MoE) strategy for the Feed-Forward Networks (FFNs) in our Transformer encoder, inspired by the DeepSeekMoE architecture~\cite{dai2024deepseekmoe}. Different from traditional MoE architectures, such as GShard~\cite{lepikhin2020gshard}, DeepSeekMoE utilizes finer-grained experts, designates some as \textit{shared experts} and the rest as \textit{routed experts}
This architectural design is well-suited to capture domain-specific features while maintaining high performance and computational efficiency.
For the FFN input $u_t$ of the t-th token, the computation of the FFN hidden output $h_t$ can be formulated as follow:
\begin{equation}
    h_t = u_t + \sum_{i=1}^{N_s} FFN_i^s (u_t) + \sum_{i=1}^{N_r} g_{i,t} FFN_i^r (u_t) 
\end{equation}
\begin{equation}
    g_{i,t} = \frac{g'_{i,t}}{\sum_{j=1}^{N_r}g'_{j,t}}
\end{equation}
\begin{equation}
    g'_{i,t} = \left\{
    \begin{aligned}
        s_{i,t} &,& s_{i,t} \in Topk({s_{j,t}|1\leq j \leq N_r},K_r) \\
        0 &,& otherwise
    \end{aligned}
    \right.
\end{equation}
\begin{equation}
    s_{i,t} = Sigmoid(u_t^T e_i)
\end{equation}

\noindent where $N_s$ and $N_r$ denote the numbers of shared experts and routed experts, respectively. $FFN^s_i(\cdot)$ and $FFN^r_i(\cdot)$ demote the i-th shared expert and the i-th routed expert, respectively. $g(i,t)$ is the gating value for the i-th expert. $K_r$ is the number of activated routed experts. $s{i,t}$ is the token-to-expert affinity. $e_i$ is the centroid vector of the i-th routed expert, and $Topk(\cdot, K)$ denotes the set comprising $K$ highest scores among the affinity scores calculated for the t-th token and all routed experts.
Considering the trade-off between computational cost and performance on all three audio domains, we set $N_s=1$, $N_r=3$, and $K_r=1$. 

\subsection{Semantic Training Stage}
\label{subsec:semantic_stage}
To simultaneously enhance semantic representation capabilities while preserving high reconstruction ability, we introduce a domain-agnostic masked modeling approach for UniCodec, inspired by Wav2Vec 2.0~\cite{baevski2020wav2vec}. Notably, our approach does not add any extra modules.
Specifically, we mask a proportion of the features output from the convolution layers in the encoder before passing them into the contextual Transformer layers. Following the masking strategy of Wav2Vec 2.0~\cite{baevski2020wav2vec}, we randomly sample a proportion $p$ of all time steps to serve as starting indices and then mask the subsequent $M$ consecutive time steps from each sampled index, allowing overlapping spans.

After the contextual Transformer layers and the quantizer, the quantized output $q_t$, centered over the masked time step $t$, requires the model to identify the unmasked convolutional latent representation $c_t$ from a set of $K + 1$ convolutional latent representations $\hat{c} \in C_t$, which includes $c_t$ and $K$ distractors~\cite{gutmann2010noise,oord2018representation}.
These distractors are uniformly sampled from other masked time steps within the same utterance.
The contrastive loss is computed as:

\begin{equation}
    L_m = - log \frac{exp(sim(q_t,c_t)/K)}{\sum_{\hat{c} \in C_t} exp(sim(q_t,\hat{c})/K)}
\end{equation}

\noindent where we compute the cosine similarity $sim(a,b) = a^Tb/||a||||b||$ between quantized tokens and unmasked convolutional latent representations~\cite{he2020momentum,chen2020simple}.

Our preliminary experiments show that training from scratch with reconstruction, masked modeling, and contrastive loss is challenging, as the single-quantizer codec struggles to simultaneously perform reconstruction and mask prediction. Therefore, we first train the codec model with reconstruction-related loss following Wavtokenizer in the \textbf{initial acoustic training stage}, omitting the masking strategy. Then we introduce this \textbf{semantic training stage} with a more difficult mask prediction goal, allowing the codec to encapsulate high-level semantic information after acquiring initial reconstruction ability.

\section{Experimental Setup}
\noindent \textbf{Datasets.}
We train UniCodec on approximately 80,000 hours of data spanning speech, music, and audio domains. For the speech domain, we use Librilight~\cite{kahn2020libri}, LibriTTS~\cite{zen2019libritts}, VCTK~\cite{vctk}, and CommonVoice~\cite{ardila2019common}. For the music domain, we use Jamendo~\cite{jamendo} and MusicDB~\cite{rafii2017musdb18} datasets. For the audio domain, we use AudioSet~\cite{audioset}.
We evaluate the speech reconstruction performance on LibriTTS test-clean. We evaluate the audio and music reconstruction performance on the AudioSet eval and MusicDB test sets, respectively.

\noindent \textbf{Training details.}
Throughout the entire training process, all input speech, music, and audio samples are resampled to 24 kHz. The batch size is $10\times32$ on 32 NVIDIA A800 80G GPUs. We uniformly truncate excessively long segments in the training data to a fixed duration of 10 seconds and feed them into the model. We use the AdamW optimizer~\cite{adam-iclr2015,adamw-Ilya-frank-iclr2019} with an initial learning rate of 2e-4 and betas set to (0.9, 0.999). The learning rate is decayed based on a cosine scheduler~\cite{cosineannealing-Ilya-frank-iclr2017}. 

During training, we provide a domain ID for each sample to allow the model to use partitioned domain-adaptive codebook to capture the distinct characteristics of each domain. However, for fair comparisons during evaluation, we do not provide domain IDs; instead, we rely on the codebook to autonomously learn the distinct paradigms of each domain and rely on the quantizer to select the nearest token from the entire codebook. As explained in Section~\ref{sec:method}, we design initial acoustic training and semantic training stages for UniCodec to balance acoustic and semantic capabilities. We follow the Wav2vec 2.0~\cite{baevski2020wav2vec} mask strategy and configuration. The mask ratio $p$ and mask length $M$ is set to 0.1 and 5.

\begin{table*}[!t]
    \centering
    \small
   \caption{\textbf{Objective reconstruction results} of UniCodec and baselines on \textbf{speech, music and audio} domains on LibriTTS test-clean, MusicDB test set, and Audioset eval set, in terms of Mel Distance and STFT Distance.
   \textbf{TPS} denotes token per second. We \textbf{bold} the best results in all the models, and \underline{\textbf{bold and underline}} the best results in single-codebook codec models.} 
   \resizebox{\textwidth}{!}{%
    \begin{tabular}{c|cc|cc|cc|cc}
      \toprule
      \multirow{2}{*}{\textbf{Model}} & \multirow{2}{*}{\textbf{Unified}} & \multirow{2}{*}{\textbf{TPS$\downarrow$}} & \multicolumn{2}{c|}{\textbf{LibriTTS test-clean}} & \multicolumn{2}{c|}{\textbf{MusicDB test}} & \multicolumn{2}{c}{\textbf{AudioSet eval}} \\
      \cmidrule(lr){4-5} \cmidrule(lr){6-7} \cmidrule(lr){8-9} 
      &&&Mel Dist$\downarrow$ & STFT Dist$\downarrow$ & Mel Dist$\downarrow$ & STFT Dist$\downarrow$ & Mel Dist$\downarrow$ & STFT Dist$\downarrow$ \\
      \toprule
      DAC & $\usym{2714}$ & 600 & 0.3697 & 1.5525 & \textbf{0.3578} & \textbf{1.9621} & 0.4581 & 2.1378 \\
      Encodec & $\usym{2714}$ & 600 & 0.5367 & 1.8271 &0.5565 & 2.1678 & 0.7601 & 2.6273 \\
      Mimi & $\usym{2714}$  & 100 & 0.6709 & 1.9859 & 0.6714 & 2.2526 & 0.8406 & 2.6639\\
      TAAE & $\usym{2718}$ & 50 & 0.7508 & 2.2426 & 1.4067  & 4.1340  & 1.9335 & 5.2897 \\
      DAC & $\usym{2718}$ & 75 & 0.7217  & 2.1662 &1.8894 & 6.2476& 1.7063 & 5.2923 \\
      BigCodec & $\usym{2718}$ & 80 & 0.4427 & 1.7385 & 1.3803  & 4.2366 & 1.8632 & 5.6171 \\ 
      Wavtokenizer~(speech) & $\usym{2718}$ & 75 & 0.5001 & 1.7879 & 0.6586 & 3.0335 & 0.5990 & 2.5479 \\
      Wavtokenizer~(music/audio) & $\usym{2718}$ & 75 & 0.5451 & 1.8649 & 0.4516 & 2.2450 & 0.4536 & 2.1871 \\
      Wavtokenizer~(unified) & $\usym{2714}$ & 75 & 0.5308 & 1.8614 & 0.5435 & 2.5451 & 0.5193 & 2.3727 \\
      UniCodec \textbf{(Ours)} & $\usym{2714}$ & 75 & \textbf{0.3442} & \textbf{1.5147} & \underline{\textbf{0.3959}} & \underline{\textbf{2.1822}} & \textbf{0.3820} & \textbf{2.1065} \\
      \bottomrule
    \end{tabular}
    }
    \label{tab:unified}
\end{table*}

Training with large-scale and diverse dataset in both acoustic and semantic stages ensure generalization ability of UniCodec. However, our preliminary experiments indicate that large-scale data training performs worse compared to training on only LibriTTS dataset. Upon analysis, we find that diverse and noisy data significantly hinders codec reconstruction learning. 
To further improve the reconstruction ability, we select high-quality data for a further \textbf{fine-tuning stage}. More details about the fine-tuning stage are in Appendix~\ref{sec:finetune}.

\noindent \textbf{Evaluation Metrics.} We adopt a comprehensive set of evaluation metrics, as follows.

    \noindent \textbf{Tokens Per Frame (TPF):} The number of parallel tokens per timestep of encoded audio, affecting ease of modeling token sequences in generative models.
    
    \noindent \textbf{Tokens Per Second (TPS):} The number of tokens per second. It determines the context length required by a generative model, especially when residual tokens are used in flattened form.
    
    \noindent \textbf{Downsample Rate (DR):} The token compression rate. It is calculated by dividing the input audio sample rate by TPS, indicating the difficulty of compressing audio waveforms into tokens.
    
    \noindent \textbf{Mel Distance (Reconstruction):} L1 distance between the mel-scaled magnitude spectrograms of the ground truth and the generated sample.
    
    \noindent \textbf{STFT Distance (Reconstruction):} L1 distance between time-frequency representations of the ground truth and the prediction, computed using multi-scale Short-Time Fourier Transform (STFT).

    More details about the metrics for speech reconstruction evaluation can be found in Appendix~\ref{sec:speechmetric}.

\noindent \textbf{Baselines.}
We select both state-of-the-art (SOTA) \textit{multi-layer} quantizer codec models and \textit{single-layer} quantizer codec models as the baselines. For multi-layer codecs, we compare against DAC~\cite{nips2023-dac-rvqgan}, Encodec~\cite{tmlr2023-encodec}, SpeechTokenizer~\cite{zhang2023speechtokenizer}, and Mimi~\cite{defossez2024moshi}. For single-layer codecs, we compare with the official checkpoints provided by Wavtokenizer~(speech)~\footnote{wavtokenizer\_medium\_speech\_320\_24k\_v2.ckpt}, Wavtokenizer~(music and audio)~\footnote{wavtokenizer\_medium\_music\_audio\_320\_24k\_v2.ckpt}, BigCodec~\cite{arxiv2024-bigcodec}~\footnote{huggingface.co/Alethia/BigCodec/resolve/main/bigcodec.pt}, and TAAE~\cite{arxiv2024-taae}~\footnote{huggingface.co/stabilityai/stable-codec-speech-16k}. 

\begin{table*}[!t]
    \centering
    \small
   \caption{\textbf{Objective reconstruction results} on the \textbf{Speech} domain from UniCodec and baselines on LibriTTS test-clean, in terms of naturalness, distortion, and intelligibility. \textbf{DR} denotes the Downsample Rate (the input audio sample rate division by Tokens Per Second (TPS)). \textbf{Unified} denotes the codec model can support all three domains of speech, music, and sound. The results of models marked by $^\dagger$ are cited from the Wavtokenizer paper~\cite{ji2024wavtokenizer} and others are reproduced by us based on the checkpoints released by the corresponding work.}
    \begin{tabular}{c|c|ccc|ccccc}
      \toprule
      \textbf{Model} & \textbf{Unified} & \textbf{DR}~($\uparrow$) & \textbf{TPF}~($\downarrow$) & \textbf{TPS}~($\downarrow$) & \textbf{PESQ}~($\uparrow$) & \textbf{STOI}~($\uparrow$) & \textbf{F1}~($\uparrow$) & \textbf{UTMOS}~($\uparrow$) \\
      \toprule
       Ground Truth$^\dagger$ & - & - & - & - & - & - & - & 4.0562 \\
      DAC & $\usym{2714}$ & 40 & 8 & 600 & 3.5197 & 0.9709 & 0.9546 & 3.6905 \\
      Encodec$^\dagger$ & $\usym{2714}$ & 40 & 8 & 600 & 2.7202 & 0.9391 & 0.9527 & 3.0399 \\
      SpeechTokenizer$^\dagger$ & $\usym{2718}$ & 40 & 8 & 600 & 2.6121 & 0.9165 & 0.9495 & 3.8794 \\
      Mimi & $\usym{2714}$ & 240 & 8 & 100 & 2.2695 & 0.9118 & 0.912 & 3.5731 \\
      TAAE & $\usym{2718}$ & 320 & 2 & 50 & 1.8955 & 0.8816 & 0.9260 & 4.1389 \\
      \midrule
      DAC & $\usym{2718}$ & 320 & 1 & 75 & 1.1763 & 0.7739 & 0.7560 & 1.3531 \\
      BigCodec & $\usym{2718}$ & 200 & 1 & 80 & 2.6872 & 0.9293 & 0.9480 & 4.0367 \\ 
      Wavtokenizer~(speech)$^\dagger$ & $\usym{2718}$ & 320 & 1 & 75 & 2.3730 & 0.9139 & 0.9382 & \textbf{4.0486} \\
      Wavtokenizer~(unified) & $\usym{2714}$ & 320 & 1 & 75 & 1.8379 & 0.8718 & 0.9175 & 3.6115 \\
      UniCodec \textbf{(Ours)} & $\usym{2714}$ & 320 & 1 & 75 & \textbf{3.0266} & \textbf{0.9493} & \textbf{0.9486} & 3.9873 \\
      \bottomrule
    \end{tabular}
    \label{tab:speech}
\end{table*}

\begin{table*}[!t]
    \centering
    \small
   \caption{\textbf{Subjective MUSHRA test reconstruction results} from codec models on \textbf{speech, music and audio} domains, on LibriTTS test-clean, MusicDB test set and AudioSet eval set. We report mean and standard deviation.}
    \begin{tabular}{c|c|cccc}
      \toprule
      \textbf{Model} & \textbf{Unified} & \textbf{LibriTTS test-clean}~($\uparrow$) & \textbf{MusicDB test}~($\uparrow$) & \textbf{AudioSet eval}~($\uparrow$) \\
      \midrule
      Ground Truth & - & 93.52 $\pm$ 1.99 & 96.18 $\pm$ 1.47 & 95.28 $\pm$ 2.18\\
      Wavtokenizer~(speech) & $\usym{2718}$ & 85.44 $\pm$ 2.29 & - & - \\
      Wavtokenizer~(music \& audio) & $\usym{2718}$ & - & 75.24 $\pm$ 2.38& 80.19 $\pm$ 2.43 \\
      Wavtokenizer~(unified) & $\usym{2714}$ & 80.40 $\pm$ 2.54 & 56.10 $\pm$ 3.74 & 62.21 $\pm$ 3.42\\
      UniCodec \textbf{(Ours)} & $\usym{2714}$ & \textbf{90.74} $\pm$ \textbf{2.06} & \textbf{77.77} $\pm$ \textbf{2.45} & \textbf{82.43} $\pm$ \textbf{2.56}\\
      \bottomrule
    \end{tabular}
    \label{tab:mushra}
\end{table*}

\section{Results and Discussions}
\subsection{Reconstruction Evaluation}
We compare the reconstruction performance of UniCodec against a broad selection of SOTA and competitive codec models as baselines. 
Table~\ref{tab:unified} presents the results of UniCodec and baselines on speech (LibriTTS test-clean), music (MusicDB test), and audio (AudioSet eval) domains, in terms of Mel Distance and STFT Distance.
As shown in Table~\ref{tab:unified}, UniCodec demonstrates excellent reconstruction performance on all three domains, outperforming the unified single-codebook baseline Wavtokenizer~(unified) and also speech-specific single-codec baselines such as BigCodec, TAAE, and Wavtokenizer~(speech).
In the music and audio domains, UniCodec also outperforms the music/audio-specific baseline Wavtokenizer~(music/audio) on both MusicDB test set and AudioSet eval set.
Even when compared to multi-layer RVQ-based unified baselines such as Encodec and Mimi, the single-layer unified UniCodec shows superior performance across all three domains, except for slightly lower performance compared to DAC (which has a much larger tokens-per-second rate) in the music domain.
The Real-Time Factors~(RTF) and comparisons of the number of parameters can be found in Appendix~\ref{sec:rtf}. 

Table~\ref{tab:speech} further compares the speech domain reconstruction performance of different codec models on \textbf{LibriTTS test-clean}, using PESQ, STOI, F1 and UTMOS, assessing the codecs in terms of naturalness, distortion, and intelligibility.
The unified UniCodec significantly outperforms WavTokenizer~(unified) across all metrics. 
Even compared to WavTokenizer~(speech) and BigCodec, which are SOTA speech-specific models with single-layer quantizers, UniCodec achieves better PESQ and STOI, demonstrating superior reconstruction quality. Furthermore, despite having a much higher downsampling rate~(DR), UniCodec remains competitive with multi-layer quantizer models such as Encodec, Mimi, and SpeechTokenizer, which have higher tokens per second (TPS).
Appendix~\ref{sec:testother} also reports the reconstruction performance on \textbf{LibriTTS test-other}.

The reconstruction results of the MUSHRA subjective test are shown in Table~\ref{tab:mushra}. UniCodec outperforms WavTokenizer~(unified) markedly in reconstruction quality across speech, music, and audio domains. Even when compared to domain-specific codecs, UniCodec performs slightly better than WavTokenizer~(speech) in the speech domain, and WavTokenizer~(music/audio) in the music and audio domains.
These results further demonstrate that \textbf{in all three domains, UniCodec achieves superior subjective reconstruction performance while maintaining a high compression rate}.


\begin{table*}[!t]
    \centering
    \small
   \caption{\textbf{Semantic representation evaluation results} on the ARCH benchmark,  in terms of classification accuracy. The results of models marked by $^\dagger$ ~are cited from the Wavtokenizer paper~\cite{ji2024wavtokenizer}. }
    \resizebox{\textwidth}{!}{%
    \begin{tabular}{l|c|cc|cc|cc}
      \toprule
      \multirow{2}{*}{Model} & \multirow{2}{*}{TPS~($\downarrow$)} & \multicolumn{2}{c|}{Speech} & \multicolumn{2}{c|}{Music} & \multicolumn{2}{c}{Audio} \\
      \cmidrule(lr){3-4} \cmidrule(lr){5-6} \cmidrule(lr){7-8}
      & & RAVDESS~($\uparrow$) & AM~($\uparrow$) & MTT~($\uparrow$) & MS-DB~($\uparrow$) & ESC50~($\uparrow$) & VIVAE~($\uparrow$) \\
      \midrule
      Encodec$^\dagger$ & 150 & 27.43 & 36.49 & 19.00 & 32.45 & 16.99 & 26.30 \\
      DAC$^\dagger$ & 100 & 25.00 & 62.87 & 25.02 & 51.37 & 20.65 & 29.91 \\
      Wavtokenizer~(speech)$^\dagger$ & 75 & 32.55 & 69.57 & - & -  & - & - \\
      Wavtokenizer~(music\&audio)$^\dagger$ & 75 & -  & - & 28.35 & 57.64 & 25.50 & \textbf{35.63} \\
      \textbf{UniCodec} & 75 & \textbf{40.28} & \textbf{70.94} & \textbf{29.55} & \textbf{59.29} & \textbf{26.00} & 34.17 \\
      ~~~~w/o semantic stage & 75 & 36.81 & 69.84 & 28.09 & 54.05 & 20.80 & 30.21 \\
      \bottomrule
    \end{tabular}
    }
    \label{tab:arch}
\end{table*}

\begin{table*}[!t]
    \centering
    \small
   \caption{Ablation study of UniCodec by evaluating the effects of domain ID during evaluation, the domain MoE module, domain-adaptive codebook, and the semantic training stage and the fine-tuning stage.}
    \begin{tabular}{l|cc|cc|cc}
      \toprule
      \multirow{2}{*}{\textbf{Model}} & \multicolumn{2}{c|}{\textbf{LibriTTS test-clean}} & \multicolumn{2}{c|}{\textbf{MusicDB test}} & \multicolumn{2}{c}{\textbf{AudioSet eval}} \\
      \cmidrule(lr){2-3} \cmidrule(lr){4-5} \cmidrule(lr){6-7} 
      & Mel Dist $\downarrow$ & STFT Dist $\downarrow$ & Mel Dist $\downarrow$ & STFT Dist $\downarrow$ & Mel Dist $\downarrow$ & STFT Dist $\downarrow$ \\
      \midrule
      UniCodec & \textbf{0.3442} & \textbf{1.5147} & 0.3959 & 2.1822 & \textbf{0.3820} & 2.1065 \\
      ~~~~w. domain id & 0.3474 & 1.5151 & \textbf{0.3912} & \textbf{2.1818} & 0.3824 & \textbf{2.1061} \\
      ~~~~w/o finetune stage & 0.4476 & 1.7005 & 0.4490 & 2.2505 & 0.4366 & 2.1659 \\
      ~~~~w/o semantic\&finetune stage & 0.4481 & 1.6978 & 0.4534 & 2.2690 & 0.4380 & 2.1723 \\
      ~~~~~~~~w/o MoE & 0.4883 & 1.8024 & 0.4592 & 2.3153 & 0.4548 & 2.2633\\
      ~~~~~~~~w/o partitioned codebook & 0.4873 & 1.7742 & 0.5064 & 2.3031 & 0.5135 & 2.2382 \\
      \bottomrule
    \end{tabular}
    \label{tab:ablation}
\end{table*}

\subsection{Semantic Evaluation}
We evaluate the semantic richness of different codec models on several speech, music, and audio domain datasets of the ARCH benchmark~\cite{ARCH}. The speech domain includes the RAVDESS~\cite{livingstone2018ryerson} and Audio-MNIST~\cite{becker2024audiomnist} datasets, the music domain includes the MTT~\cite{law2009evaluation} and MS-DB~\cite{rafii2017musdb18} datasets, and the audio domain includes the ESC50~\cite{piczak2015esc} and VIVAE~\cite{holz2022variably} datasets.
We extract embeddings corresponding to the discrete codebooks of each acoustic codec model as its respective representations and evaluate the classification accuracy of the codec models on the ARCH datasets using these representations. The experimental results, as shown in Table~\ref{tab:arch}, demonstrate that our UniCodec outperforms WavTokenizer, DAC (configured with a single quantizer) and Encodec (configured with two-layer quantizers), in terms of classification accuracy. Furthermore, performance comparison against the counterpart that excludes the semantic stage training (w/o semantic stage) verifies the effectiveness of the proposed semantic training using mask prediction and contrastive loss.
In future work, we plan to explore UniCodec-based ALM on downstream audio tasks such as audio continuation and generation.

\subsection{Ablation study}
\label{subsec:ablation}
We conduct ablation study by evaluating the effect of proposed methods and modules on the LibriTTS test-clean, MusicDB test, and AudioSet eval sets. As shown in Table~\ref{tab:ablation}, providing the domain ID for the partitioned domain-adaptive codebook during evaluation performs comparably to the default setting without providing domain ID. The only exception is the music domain, where performance improves slightly due to the inherent mixed nature of songs, which contain both speech and music elements. These results demonstrate that the partitioned domain-adaptive codebook can autonomously capture distinct domain-specific features.
The third row shows that without the fine-tuning stage, a significant performance degradation is observed when trained on large but noisy data. This highlights the critical role of high-quality data in codec training.
The fourth row reports results without both semantic training and fine-tuning stages. Comparison between the third and fourth rows shows that our proposed semantic stage enhances semantic information while preserving reconstruction ability.
Furthermore, removing the MoE module from UniCodec without the semantic and fine-tuning stages (i.e., only the initial acoustic training stage) results in an additional performance degradation. Removing the partitioned domain-adaptive codebook~(i.e. naive single codebook) leads to even greater degradation than removing the MoE module. These results confirm the effectiveness of the proposed domain MoE and partitioned domain-adaptive codebook strategy in achieving a unified codec with superior reconstruction ability.


\section{Conclusions}
In this work, we introduce UniCodec, a low-bitrate unified audio tokenizer designed to support multi-domain audio data, including speech, music, and sound, using a single quantizer. To achieve this goal of unified modeling, we propose the partitioned domain-adaptive codebook and the domain MoE strategy to capture the distinct characteristics of each domain. To enrich the semantic information without introducing additional modules, we propose a self-supervised mask prediction modeling algorithm during codec training. Comprehensive objective and subjective evaluations demonstrate that, as a unified audio codec with a single codebook, UniCodec achieves excellent performance in both acoustic and semantic capabilities.

\section{Limitations}
Our experiments reveal that UniCodec training will be disrupted by noisy or low-quality inputs. Modeling speech in complex environments, such as noisy settings or with overlapped speech, remains a challenge. We anticipate that future work will address these issues, improving model robustness for such scenarios.

Although our experiments demonstrate that the proposed semantic training stage with mask prediction and contrastive loss effectively captures semantic information, it remains challenging for a unified single-codebook codec to balance both acoustic and semantic density across diverse domain data. We believe that it is a promising research direction to focus on enhancing semantic capabilities while preserving reconstruction performance, without introducing additional modules.

We have evaluated the model in streaming use cases but have observed some performance degradation. Future work should aim to improve streaming capabilities while maintaining high reconstruction quality.

Due to space limit and computational constraints, we have focused on demonstrating UniCodec's reconstruction capabilities and have not yet explored training UniCodec with LLM to function as an Audio Language Model~(ALM). In future work, we plan to investigate the performance of UniCodec-based ALM on downstream audio tasks.

\bibliography{custom}

\newpage
\appendix
\section{Speech Reconstruction Evaluation}
\label{sec:testother}
We further evaluate UniCodec on the LibriTTS test-other set to assess its reconstruction ability on noisy data. The results in Table~\ref{tab:testother} show that the reconstructed speech from our model achieves a higher UTMOS score than the ground truth on the LibriTTS test-other noisy dataset. This indicates that UniCodec reconstructs speech with greater naturalness and quality, even in the presence of noise. As a unified codec with a single codebook, UniCodec outperforms Wavtokenizer~(unified) across all metrics. Even when compared with other state-of-the-art speech-specific codecs with a single codebook, UniCodec maintains competitive performance.

\begin{table*}[!t]
    \centering
    \small
   \caption{\textbf{Objective reconstruction results} on the \textbf{Speech} domain from UniCodec and baselines on LibriTTS test-other, in terms of naturalness, distortion, and intelligibility. \textbf{DR} denotes the Downsample Rate (the input audio sample rate division by Tokens Per Second (TPS)). \textbf{Unified} denotes the codec model can support all three domains of speech, music, and sound. The results of models marked by $^\dagger$ are cited from the Wavtokenizer paper~\cite{ji2024wavtokenizer} and others are reproduced by us based on the checkpoints released by the corresponding work.}
    \begin{tabular}{c|c|ccc|cccc}
      \toprule
      \textbf{Model} & \textbf{Unified} & \textbf{DR}~($\uparrow$) & \textbf{TPF}~($\downarrow$) & \textbf{TPS}~($\downarrow$) & \textbf{PESQ}~($\uparrow$) & \textbf{STOI}~($\uparrow$) & \textbf{F1}~($\uparrow$) & \textbf{UTMOS}~($\uparrow$) \\
      \toprule
      Ground Truth$^\dagger$ & - & - & - & - & - & - & - & 3.4831  \\ 
      DAC$^\dagger$ & $\usym{2714}$ & 48.9 & 9 & 900 & 3.7595 & 0.9576 & 0.9696 & 3.3566 \\
      Encodec$^\dagger$ & $\usym{2714}$ & 40 & 8 & 600 & 2.6818 & 0.9241 & 0.9338 & 2.6568 \\
      SpeechTokenizer$^\dagger$ & $\usym{2718}$ & 40 & 8 & 600 & 2.3269 & 0.8811 & 0.9205 & 3.2851 \\
      Mimi & $\usym{2714}$ & 240 & 8 & 100 & 2.0952 & 0.8816 & 0.8875 & 3.0608 \\
      TAAE & $\usym{2718}$ & 320 & 2 & 50 & 1.7539 & 0.8380 & 0.8994 & 3.7136 \\
      \midrule
      DAC$^\dagger$ & $\usym{2718}$ & 440 & 1 &100 & 1.2454 & 0.7505 & 0.7775 & 1.4986 \\
      BigCodec & $\usym{2718}$ & 200 & 1 & 80 & \textbf{2.3817}  & 0.9094  & \textbf{0.9237} & 3.5453 \\ 
      Wavtokenizer~(speech)$^\dagger$ & $\usym{2718}$ & 320 & 1 & 75 & 2.2614 & 0.8907 & 0.9172 & 3.4312 \\
      Wavtokenizer~(unified) & $\usym{2714}$ & 320 & 1 & 75 & 1.6649 & 0.8312 & 0.8874 & 3.0820 \\
      UniCodec & $\usym{2714}$ & 320 & 1 & 75 & 2.2749 & \textbf{0.9095} & 0.9109 & \textbf{3.5800} \\
        \bottomrule
    \end{tabular}
    \label{tab:testother}
\end{table*}

\section{Real-Time Factor}
\label{sec:rtf}
To evaluate the real-time performance of different audio codec models, we compute the Real-Time Factor (RTF) for audio durations of 5, 10, 30, and 60 seconds. The evaluation is conducted on a test set of 1,000 audio clips to ensure a robust and fair comparison. All experiments are performed on an NVIDIA A100 GPU. RTF measures the processing speed relative to real-time feature extraction, a critical metric for NACs to minimize latency. Lower RTF values indicate faster processing.
As shown in Table~\ref{tab:rtf}, UniCodec has more parameters than Wavtokenizer due to the incorporation of transformer layers and the MoE structure. This results in a higher RTF for UniCodec with 5-second inputs compared to Wavtokenizer. However, for 10, 30, and 60-second inputs, UniCodec exhibits better RTF performance, and benefits from the superior parallel processing capabilities of its transformer layers, compared to the LSTM module in Wavtokenizer.
Semanticodec has a much larger RTF, making it unsuitable for real-time applications. For DAC, we do not report results for 30s and 60s due to out-of-memory issues.

\begin{table*}[!t]
    \centering
   \caption{Real-Time Factors~(RTFs) for audio codec models on test audio clips of 5s, 10s, 30s and 60s duration using an A100 GPU.}
    \begin{tabular}{c|ccccc}
      \toprule
      Model & Parameter~(M) & RTF~(5s)$\downarrow$ & RTF~(10s)$\downarrow$ & RTF~(30s)$\downarrow$ & RTF~(60s)$\downarrow$ \\
      \midrule
      DAC & 76 &0.01021 &0.00771& - & - \\
      SemantiCodec & 507 & 1.10905 &0.54455 & 0.69320 & 0.61164\\
      Wavtokenizer & 77 & \textbf{0.00377} & 0.00321 & 0.00286 & 0.00280 \\
      UniCodec & 274 & 0.00467 & \textbf{0.00287} & \textbf{0.00196} & \textbf{0.00187}\\
      \bottomrule
    \end{tabular}
    \label{tab:rtf}
\end{table*}

\section{Fine-tuning Stage}
\label{sec:finetune}

In the finetune stage, we select high-quality speech data with a high UTMOS, including LibriTTS train-clean, VCTK, and LJSpeech~\cite{ljspeech17}. Additionally, the learning rate and mel loss coefficient are set to 5e-5 and 450, respectively. These training strategies in the finetune stage significantly enhance the model’s ability to better learn reconstruction ability.

\section{Codebook Utilization}
\label{sec:codebook}
We further evaluate the codebook utilization rate for both the entire codebook and the partitioned codebook across each domain. The results are evaluated on the LibriTTS test-clean, MusicDB test, and AudioSet eval sets. As shown in Table~\ref{tab:codebook}, the utilization rates for each domain-partitioned codebook are nearly fully exploited, demonstrating that our UniCodec's domain-adaptive codebook is both well-trained and effectively utilized.

\begin{table}[!t]
    \centering
    \small
   \caption{Codebook utilization rate of the whole codebook and three domain-partitioned codebook in the condition of with and without domain id provided.}
    \begin{tabular}{c|ccccc}
      \toprule
      & Whole & Speech & Music & Audio \\
      \midrule
      w/o domain id & 99.63\% & 98.54\% &100\% & 99.95\% \\
      w. domain id & 99.62\% & 98.54\% &100\% & 99.96\% \\
      \bottomrule
    \end{tabular}
    \label{tab:codebook}
\end{table}

\section{Speech Reconstruction Metrics}
\label{sec:speechmetric}
\noindent \textbf{PESQ~\cite{rix2001perceptual} (Distortion):} A speech quality assessment metric that compares reconstructed speech with reference speech, with scores ranging from 1 to 5, and correlates with human judgment.
    
\noindent \textbf{STOI (Intelligibility):} A metric measuring speech intelligibility by comparing short-time spectral envelopes between reconstructed and ground truth speech, with scores ranging from 0 to 1.
    
\noindent \textbf{F1 Score (Voiced/Unvoiced Classification):} It balances precision and recall for voiced/unvoiced classification.
    
\noindent \textbf{UTMOS~\cite{saeki2022utmos} (Naturalness):} An automatic speech MOS (Mean Opinion Score) predictor evaluates the naturalness of generated speech, reflecting overall auditory quality.

\end{document}